\documentstyle[12pt,epsfig,epsf]{article}
\begin{document}

\begin{flushright}
Freiburg--THEP 96/15\\
August 1996
\end{flushright}

\vspace{1.5cm}

\begin {center}

{\large \bf Influence of strongly coupled, hidden scalars \\
            on Higgs signals}
            
\vskip 1cm

{\bf T. Binoth, J. J. van der Bij }\\ 

\bigskip 

Albert--Ludwigs--Universit\"at Freiburg\\
Fakult\"at f\"ur Physik \\ Hermann--Herder--Stra\ss e 3\\
79104 Freiburg i. Br.\\
\end{center}

\vspace{3cm}

\abstract{\em 
To investigate the possible effects of a light hidden sector on 
Higgs boson detection, we discuss a model of scalar singlets coupled to the 
Standard Model. The model effectively makes the Higgs width a free parameter due
to additional invisible decay modes. This width can become arbitrarily large.
Theoretical and experimental bounds on model parameters are presented.
It is shown, how Standard Model predictions change and that in the 
case of large coupling, Higgs signals will be diluted. We study, to which
extent such a strongly coupled, hidden sector can be excluded by present
and future Higgs search experiments.   }

\newpage
\section{Introduction}
A major task of future high energy collider experiments is the search 
for the Higgs boson. Its
detection would be the ultimate confirmation of the spontaneous symmetry
breaking mechanism. In the popular standard scenarios of particle
physics, the Standard Model (SM) and its minimal supersymmetric generalization
(MSSM), lower bounds on the Higgs boson of the SM and the lightest CP--even
Higgs scalar of the MSSM exist due to the LEP1 experiment \cite{opal}.
If no events are found  by LEP2, the SM Higgs bounds will be increased up to
a mass of about ($\sqrt{s} - 100)$ GeV with $\sqrt{s}$ somewhere
between 175 and 205 GeV \cite{LEP2report}. In
the SM and MSSM the very small width of the Higgs boson, a few MeV, leads
to sharply peaked resonances, if one plots the cross section
as a function of the recoiling mass of the Higgs decay products \cite{riemann}. 
The signal to background ratio in the resonance region
is almost entirely determined by the experimental energy resolution, though
initial state QED corrections play a role there, too. Thus, the detection of 
the Higgs boson inside these models is only a question of its mass. 

In this paper we want to analyze, to which extent the Higgs mass bounds 
are affected by a light hidden sector. To this end we study a simple extension 
of the scalar sector of the SM, that can lead to a dilution of Higgs boson
signals. We stress that such a hidden sector not only influences Higgs signals 
by nonstandard invisible Higgs decay but also
by a broadening of the resonance, which considerably affects the  signal
to background ratio, $S/B$, in the case of large couplings. In order to 
discuss the consequences of the non-observation of a light Higgs we mention 
this point, because strong interactions are not usually believed
to play a role at LEP/NLC energies.
The invisible decays from Majoron or neutralino
decay discussed in the literature are leading mainly
to a modification of branching ratios and not to an
increase of the Higgs width beyond the order of the experimental resolution, a 
few GeV. 

In models, where the Higgs width is significantly enhanced, $S/B$ will 
unavoidably be reduced.
By adding matter from a hidden sector
to the SM and allowing for relatively strong couplings between the Higgs boson 
and the hidden matter, such an effect can be induced. 
To make quantitative statements
we analyze an $O(N)$--symmetric scalar model of gauge singlets which shall 
serve as a toy model for all kinds of light hidden matter physics. 
To allow for nonperturbative couplings we use $1/N$ expansion techniques.
The existence of such singlets
would not at all effect present experimental results, because they do not
couple to the known fermions. Gauge singlet fields occur already
in Majoron models \cite{valle}, MSSM plus singlet \cite{nmssm}
and in models, where the baryon asymmetry of the universe is explained
by electroweak physics \cite{bau}.

In the next section our model will be introduced. We show, how lower bounds
on the Higgs mass, due to vacuum instability and upper bounds,
due to the Landau pole of the theory, depend on our model parameters. 
Afterwards we discuss, how 
signals which should lead to the detection of a kinematically allowed Higgs 
boson are modified 
by the model, how the non-observation of the Higgs boson at LEP1 already 
restricts
the model parameters and how the LEP2 and NLC experiments will do so.   
To this end we have to compute the Higgs production cross section
including the dominant background contributions for the most prominent signals. 
The analysis results in  exclusion plots for the theoretically and 
experimentally allowed parameters. We determine the parameter region, 
where a kinematically allowed Higgs boson
would be visible at LEP/NLC. The discussion on the phenomenology of the model
will be closed by some remarks on invisible Higgs search at the LHC. 
The results are summarized resumed in a short conclusion.
 
\section{The Higgs--$O(N)$-singlet model} 
To illustrate the consequences of a hidden sector coupled to the Higgs boson
in a possibly strong way, we want to consider the case of scalar gauge singlets
-- let us call them ''Phions'' for shortness -- added to the SM. 
To deal with the  case of strong interactions
we introduce an $N$--plet
of such Phions. This allows us to use nonperturbative $1/N$--methods.
Neglecting all the fermions and gauge couplings for the moment, 
our model consists  of the
SM Higgs sector coupled to an $O(N)$--symmetric scalar model. 
Similar models can be found in Ref.~\cite{chivukula}.
Our Lagrangian density is:
\begin{eqnarray} \label{model}
L_{Scalar} &=& L_{Higgs} + L_{Phion} + L_{Interaction}   \nonumber  \\
           & & \hspace{7cm} \mbox{where}\nonumber \\
L_{Higgs}  &=& 
 - \partial_{\mu}\phi^+ \partial^{\mu}\phi -\lambda_0 \,
 (\phi^+\phi - \frac{v_0^2}{2})^2 \nonumber \\
L_{Phion}  &=& - \frac{1}{2}\,\partial_{\mu} \vec\varphi \, 
\partial^{\mu}\vec\varphi
     -\frac{1}{2} m_{P0}^2 \,\vec\varphi^2 - \frac{\kappa_0}{8N} \, 
     (\vec\varphi^2 )^2 \nonumber \\
L_{Inter.} &=& -\frac{\omega_0}{2\sqrt{N}}\, \, \vec\varphi^2 \,\phi^+\phi 
\end{eqnarray}  
Here we use a metric with signature $(-+++)$. 
$\phi=(\sigma+v+i\pi_1,\pi_2+i\pi_3)/\sqrt{2}$ is the complex Higgs doublet of 
the SM with the
vacuum expectation value $<0|\phi|0> = (v/\sqrt{2},0)$, $v=246$ GeV. Here, 
$\sigma$ is the physical  
Higgs boson and $\pi_{i=1,2,3}$ are the three Goldstone bosons. 
$\vec\varphi = (\varphi_1,\dots,\varphi_N)$ is a real vector with 
$<0|\vec\varphi|0>= \vec 0$. $v_0,\lambda_0,\omega_0,\kappa_0,m_{P0}$ are bare 
parameters.If we would allow for a non-vanishing vacuum expectation 
value for the Phions, the mass matrix
would become non-diagonal and Higgs--Phion mixings would occur. 
The lightest scalar of the gauged model
would have a reduced coupling to the vector bosons by a 
cosine of a mixing angle. We will not discuss this possibility 
further, as we are mainly interested in the effects coming from the Higgs width.
If we look at the gauged model we can choose the unitary gauge to rotate away 
the unphysical Goldstone bosons. This is gauge invariant, because in the 
following we only consider loops
of gauge singlet particles. Note that the vacuum induced mass term for the 
Phions is suppressed by a factor $1/\sqrt{N}$.

In the case of large non standard couplings $\omega$ and $\kappa$, 
loop induced operators with external Higgs and Phion fields appear and are not 
negligible. They are only suppressed by powers of $1/N$. 
As we are only interested in operators with external Higgs legs, we classify 
these in types, with (a) and without (b) internal Higgs lines. Diagrammatically:

\begin{picture}(10.,3.)
\put(3.,0.4){
\epsfxsize8.0cm
\epsfysize2.5cm
\epsffile{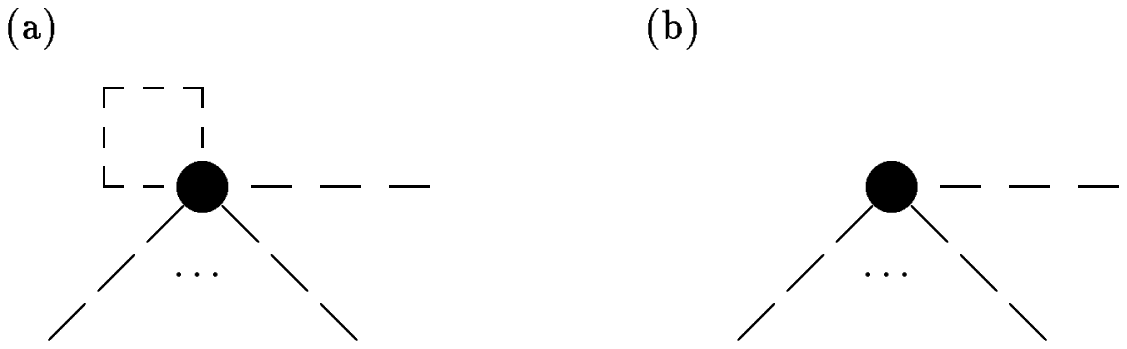}}
\end{picture}

The former (a)with $k$ legs are formed by closing two Higgs lines of a $k+2$ 
operator of type (b). 
Because operators carrying a virtual Higgs inside a loop are suppressed,
it is enough to focus on the latter. 
These we can sort with 
respect to the number of Higgs legs ($k$) and powers ($n$)
of $N$. Every Higgs--Phion vertex counts $n=-1/2$, every $\kappa$--vertex 
counts $n=-1$ and every
Phion loop counts $n=1$. The highest $n$ operators have $n=1/2$. They are

\begin{picture}(16,1.)
\put(3,0.2){
\epsfxsize5.5cm
\epsfysize0.8cm
\epsffile{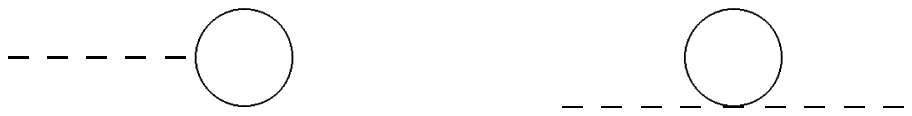}}
\end{picture}

Both contributions to the Higgs propagator lead to a trivial renormalization 
of the bare parameters for any fixed value of $N$. 
The tadpole contribution is taken
into account by using the experimental value of the vacuum expectation value 
(vev) $v = 246\,GeV$, the constant self-energy term can 
be absorbed by a bare mass term for the complex Higgs doublet. The $n=0$ 
operators form an infinite sum of loop graphs, a so called bubble sum. They are 

\begin{picture}(16,3.5)
\put(2,0.2){
\epsfxsize8.0cm
\epsfysize3.cm
\epsffile{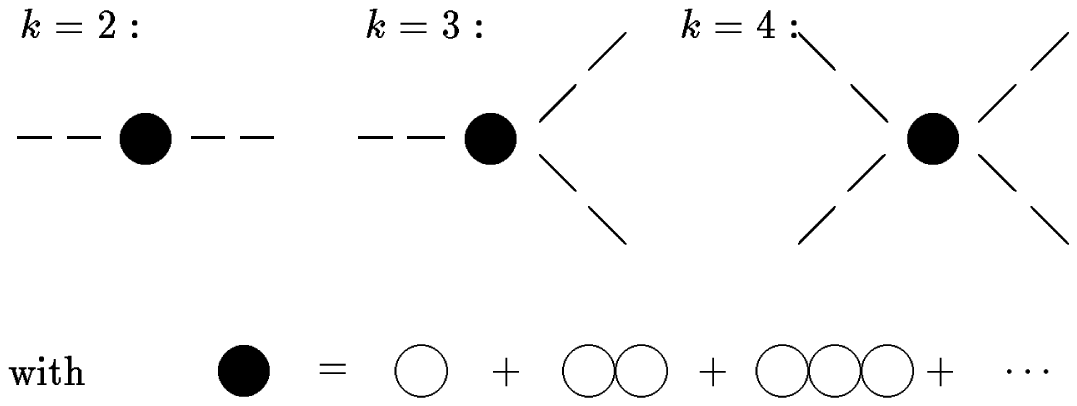}}
\end{picture}

No other structures with $n=0$ are possible. Operators with higher $n$ are 
suppressed by factors of $(1/N)^n$, which can be made arbitrarily small by 
taking $N$ sufficiently large. We only want to discuss this large--$N$ case, 
the formal limit $N\rightarrow \infty$. 
The upper $n=0,\, k=3,4$ bubble sums are leading to a renormalization of
the Higgs coupling $\lambda$, which we want to keep perturbative at the given
experimental scale.
Without this condition  a Higgs boson would not be in reach for LEP or NLC,
as long as $\sqrt{s}\leq 500$ GeV. 
 
For the discussion of Higgs signatures, it is enough to focus on the 
Higgs-propagator.The Higgs self--coupling will not play a role for Higgs search.
As shown above the propagator is modified by the Phions.
In the leading order in $1/N$, which is found in the limit $N \rightarrow 
\infty$, the Higgs self-energy is given by an infinite sum of Phion bubble 
terms. Regularization of the divergent bubbles, i. e. absorbing the divergent 
and some constant
contributions into the bare parameters, is done by subtraction of the 
logarithmically divergent part  \cite{ein}. 
With this regularization, the Euclidean bubble integral
\begin{displaymath} 
       I_{Bubble}(s=-p^2,m_P^2) = \frac{1}{2} \int \frac{d^4k}{(2\pi)^4}\,
                 \frac{1}{k^2+m_P^2}\,\frac{1}{(k+p)^2+m_P^2}  \nonumber,
\end{displaymath}
becomes above the Phion threshold
\begin{eqnarray}
 && I(s,\mu^2,m_P^2) \quad = \quad  I_{Bubble}(s,m_P^2) - I_{Bubble}(0,\mu^2)\\
    \qquad           && = -\frac{1}{32\pi^2}  \left( \log(\frac{m_P^2}{\mu^2}) 
    - 2 +  \sqrt{1-\frac{4m_P^2}{s}} 
           \left( \log \left(
           \frac{1+\sqrt{1-\frac{4m_P^2}{s}}}{1-\sqrt{1-\frac{4m_P^2}{s}}}
           \right)  - i\pi\right)\right)\nonumber 
\end{eqnarray}
with the arbitrary renormalization scale $\mu$. 
In the case of massless Phions this simply reduces to  
$I(s,\mu^2,0) = -1/(32\pi^2)(\log(s/(e\mu)^2) - i\pi)$. The bubble sum is 
the geometric series of the integral times a coupling.\\
\begin{picture}(16,1.8)
\put(2,.15){
\epsfxsize9.0cm
\epsfysize1.3cm
\epsffile{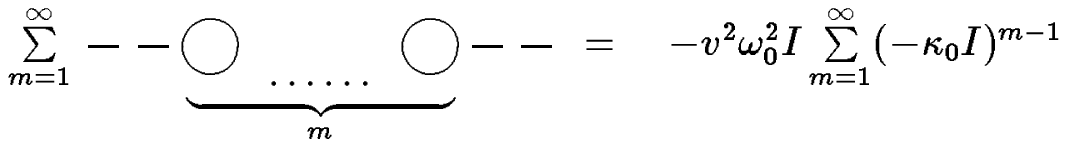} }
\end{picture}
Adding up all regularized terms gives the inverse Higgs propagator     
\begin{eqnarray} \label{propagator}
D_{H}^{-1}(s,\mu^2) &=& -s + M_H^2 - i\sqrt{s}\Gamma_{SM}(s)
                     + \Sigma (s,\mu^2)  \\
                     && \nonumber \\
\Sigma (s,\mu^2) &=&
 \frac{-\omega^2 v^2 I(s,\mu^2,m_P^2)}{1+\kappa I(s,\mu^2,m_P^2)} \nonumber
\end{eqnarray}
Below the Phion threshold the self-energy is real valued and can be absorbed 
into the renormalization of the Higgs self-coupling . 
Above the Phion threshold, $s>4\,m_P^2$, $\Sigma$
develops an imaginary part which results in a Higgs width depending on the non 
standard parameters leading to observable effects.
The independent SM Higgs-width is added, too.
To find an explicit expression for  the upper propagator, remember that inside 
the SM the Higgs mass, or better the quartic Higgs coupling,
is a free parameter. 

Defining the mass by the location of the resonance on the real $p^2$--axis 
fixes our renormalization scale $\mu$ by the equation
\begin{equation}\label{rencon}
Re( \Sigma (M_H^2,\mu^2) ) = 0
\end{equation}
Using this relation, the abbreviations  
$\tilde\omega^2=\omega^2/(32\pi^2)$, 
$\tilde\kappa=\kappa/(32\pi^2)$ and $r(x) = \sqrt{1 - 4\,m_P^2/x}$,
one finds, after splitting the integral in its real and imaginary part
\begin{eqnarray} 
I(s,\mu^2,m_P^2)\vert_{\mu\,fixed} &=& a(s)+i\,b(s) \nonumber\\
a(s) &=&  (\sqrt{1-(2\pi\tilde\kappa r(M_H^2))^2}-1)/(2\tilde\kappa) \nonumber\\
  &&  + r(M_H^2) \log(\frac{1+r(M_H^2)}{1-r(M_H^2)})- r(s) \log(\frac{1+r(s)}{1-r(s)})\nonumber\\
b(s) &=&  \pi\,r(s)  \, , \nonumber
\end{eqnarray}  
an expression for the Higgs propagator, in terms 
of running quantities:
\begin{eqnarray} 
D_{H}^{-1}(s) &=& -s + M_H(s)^2 - i\sqrt{s}\,\Gamma_{H}(s) \\ \nonumber\\
M_H(s)^2 &=& M_H^2 - \tilde\omega^2v^2  
\frac{a(s) + \tilde\kappa (a(s)^2+b(s)^2)}{(1+\tilde\kappa\,a(s))^2+
(\tilde\kappa\,b(s))^2} \nonumber\\
\Gamma_{H}(s) &=& \Gamma_{SM}(s)+\frac{\tilde\omega^2v^2}{\sqrt{s}}
\frac{b(s)}{(1+\tilde\kappa a(s))^2+(\tilde\kappa\,b(s))^2}\nonumber
\end{eqnarray}  
Remember that this expression is only valid above the Phion threshold. 

\begin{figure}[htb]
\vspace{0.1cm}
\centerline{\epsfig{figure=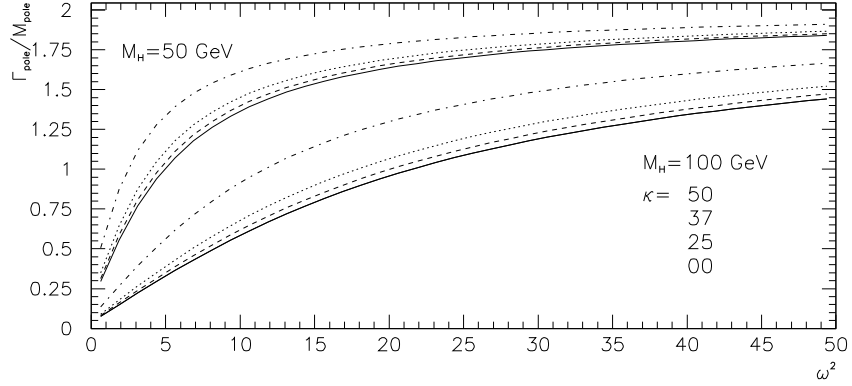,height=5.cm,angle=0}}
\begin{center} 
\parbox{13cm}{
\caption{\label{womfig}\it The ratio $\gamma_H/m_H$ for several 
values of $\kappa$ and $m_H=50, 100$ GeV. Lower lines for given $M_H$
belong to lower values of $\kappa$.}}
\end{center}
\end{figure}

For the definition of mass in the case of a large width a comment is in order.
If one defines the pole mass and width of the propagator, Eqn.~\ref{propagator},
by its zero in the complex energy plane, $p^2=-(m_H - i \gamma_H/2)^2$, 
as was done in \cite{ein},
one finds a difference to the mass and width defined by the location and width
of the resonance which was used above. The two descriptions are related to each 
other by the following
formulae
\begin{eqnarray} \label{wom}
m_H^2 = M_H^2\,(1+\sqrt{1 + \Gamma_H^2/M_H^2)})/2 &,& 
 \gamma_H^2 = 2\,\Gamma_H^2/(1+\sqrt{1 + \Gamma_H^2/M_H^2}) \nonumber \\
\Leftrightarrow \hspace{2.5cm} M_H^2 = m_H^2 - \gamma_H^2/4 &,
        & \Gamma_H^2 = \gamma_H^2/(1 - \gamma_H^2/(4 m_H^2)) \nonumber \\
\Rightarrow \hspace{4.8cm} \frac{\Gamma_H}{M_H} &=& \frac{\gamma_H}{m_H}
 \frac{1}{1-\gamma_H^2/(4 m_H^2)} 
\end{eqnarray} 
As long as $\Gamma<<M$ the two descriptions differ only slightly, because it is 
actually a higher loop effect. For small couplings the width is growing 
linearly with the squared 
coupling, and there is no significant deviation between the two definitions of 
mass and width. To 
illustrate the difference of the two mass and width descriptions we give a 
table of some limiting values
neglecting the Phion mass and self-coupling and SM width.
\begin{center}
\begin{tabular}{|c|c|c|c|c|} \hline 
$\omega$  &  $m_H/v$  &  $\gamma_H/v$  &  $\gamma_H/m_H$  &  $\Gamma_H/M_H$ \\ 
\hline
$\rightarrow \, 0$                         & 
$ \rightarrow\sqrt{2\lambda}$              & 
$\sim \omega^2/(32\pi^2\sqrt{2\lambda})$  &  
$\rightarrow \, 0$                        &
$\rightarrow \, 0$                     \\
$\rightarrow\infty$                      & 
$\sim \omega/(8\pi)$                     & 
$\sim \omega/(4\pi)$                     & 
$\rightarrow 2$                          & 
$\rightarrow \infty$       \\  \hline  
\end{tabular}
\end{center}
The width to mass ratio
of the resonance, which is the experimentally important quantity, 
is not bounded as $\omega$
is increasing. For $\omega \rightarrow \infty$ we have
$\gamma_H/m_H\rightarrow 2$ and $\Gamma_H/M_H\rightarrow \infty$,
which proves that the Higgs resonance is generically smeared out, 
if there exist a strong decay channel into light 
hidden matter. 
The dependence of the width to mass ratio is demonstrated  in 
Fig.~(\ref{womfig}), 
where we have plotted this ratio versus the Higgs Phion coupling for
two values of $M_H$ and several values of $\kappa$. The dependence
on $\kappa$ is only mild as long as $\kappa/(16\pi^2)\sim O(0.1)$. 
A further increase
of $\kappa$ would imply the presence of a Landau-pole already in the 
scalar sector. 

\begin{figure}[htb]
\vspace{0.1cm}
\centerline{\epsfig{figure=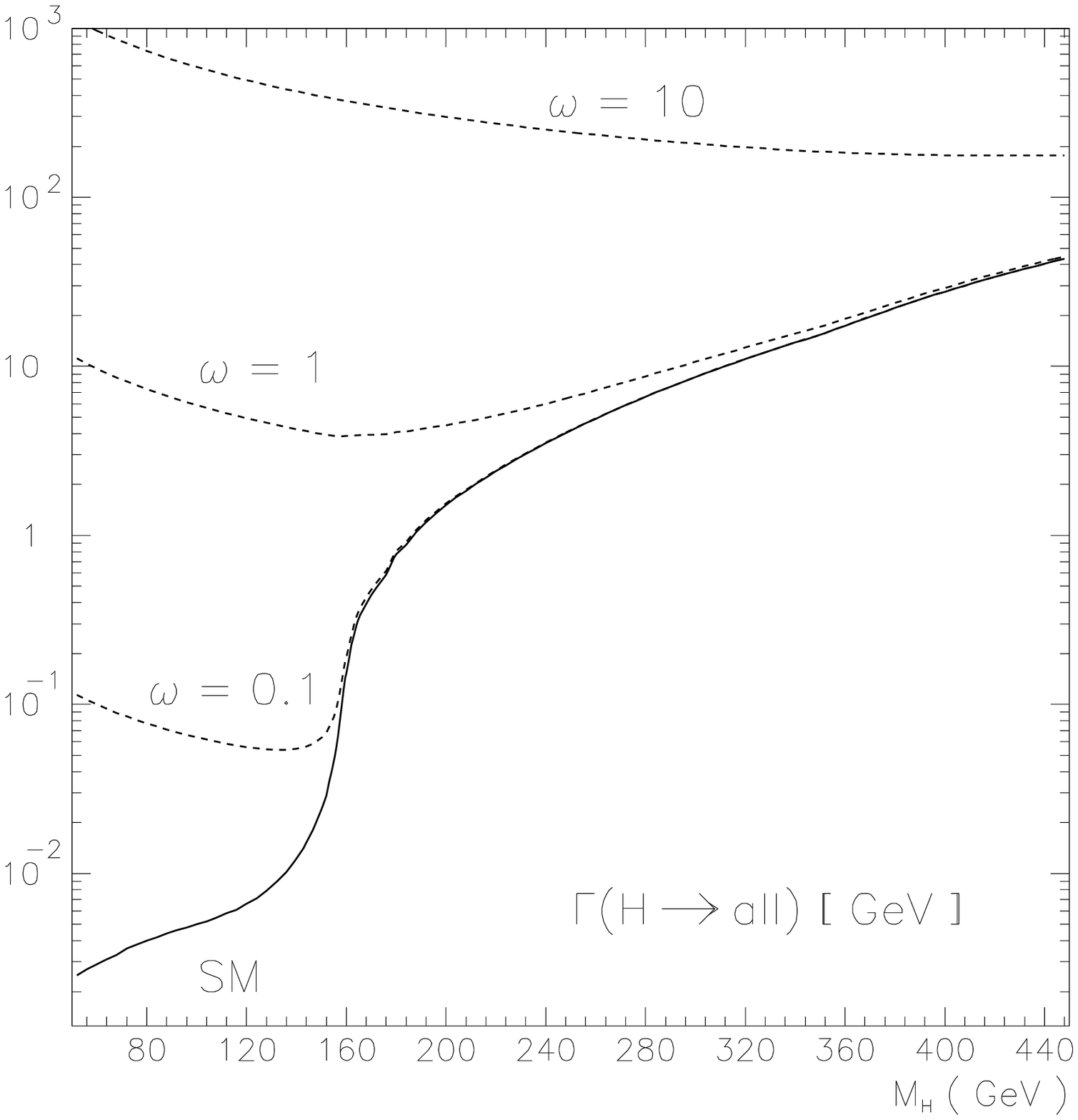,height=10.cm,angle=0}}
\begin{center}
\parbox{13cm}{
\caption{\label{width}\it Higgs width in comparison with the Standard Model.}}
\end{center}
\end{figure}

The crucial point of the model is that it makes the Higgs width
essentially a free parameter because invisible decay modes are
present. While in principle arbitrary couplings are possible 
the model allows very wide resonances which means extremely fast Higgs decay.  
The comparison of the Higgs width of the Phion model to the SM width
is shown in Fig.~(\ref{width}).

\section{Theoretical bounds} 
  
Before entering the discussion on experimental bounds we want to comment on 
theoretical restrictions on the model parameters. To this end we analyze the 
one--loop renormalization group equations (RGEs) of our model, 
because strongly interacting theories are
usually contaminated by Landau pole singularities. On the other hand the 
vacuum instability of the Higgs sector terminates the validity of the model 
at some scale \cite{sher,quiros}.
Avoiding Landau poles and vacuum instabilities below a given scale $\Lambda$, 
we find theoretical bounds on our model parameters defined at a reference 
energy $\mu_{ref.}$. To this end we calculate the one--loop  RGEs for our model 
including the top yukawa coupling
$\lambda_T$ and the gauge couplings. Together with the well known RGEs 
\cite{hill} 
from the SM, one finds with 
$t=\log(\mu/\mu_{ref.}),\partial _t  =\partial/\partial t$ 
in leading order in $1/N$
\begin{eqnarray} \label{rges}
(4\pi)^2\partial_t \lambda &=& 
       24\lambda^2 + 
       \lambda (12\lambda_T^2 - 9 g_2^2 - 3 g_1^2)\\
      && - 6\lambda_T^4 + (3g_1^4 + 6 g_1^2g_2^2 + 9g_2^4)/8 + \omega^2 
      \nonumber\\
 (4\pi)^2\partial_t \omega   &=& 
                           \omega \,(24\lambda + \kappa + 6\lambda_T^2) 
                           \nonumber \\
  (4\pi)^2\partial_t \kappa &=& 
                                        8\omega^2 + \kappa^2/3 \nonumber \\
   (4\pi)^2\partial_t \lambda_T &=& 
   9/2\lambda_T^3 - (8g_3^2+9/4g_2^2+17/12 g_1^2)\lambda_T \nonumber \\
   g_i^{-2}(t) &=& g_i^{-2}(0) + c_i/(8\pi^2)t\, , \quad c_i=(-41/6,19/6,7)_i, 
   \, (i=1,2,3)\nonumber
\end{eqnarray} 
The evolution of the couplings is determined if we fix initial conditions at 
$t=0$. We took
$\mu_{ref.}=2M_Z$ as initial point. We use the experimental data
$\alpha=1/128.9$, $\sin^2\theta_W=0.2322$, $\alpha_s=0.124$, 
$m_T=175\,GeV$ \cite{expdata}. One gets 
$\lambda_T(0)=m_T\sqrt{2}/v=1.006$, $g_1(0)^2=0.212$, $g_2(0)^2=0.420$, 
$g_3(0)^2=1.558$.
To find an exclusion plot in the $(\omega,M_H)$ plane, we vary the respective 
couplings for a fixed value of $\kappa$.

\begin{figure}[htb]
\vspace{0.1cm}
\centerline{\epsfig{figure=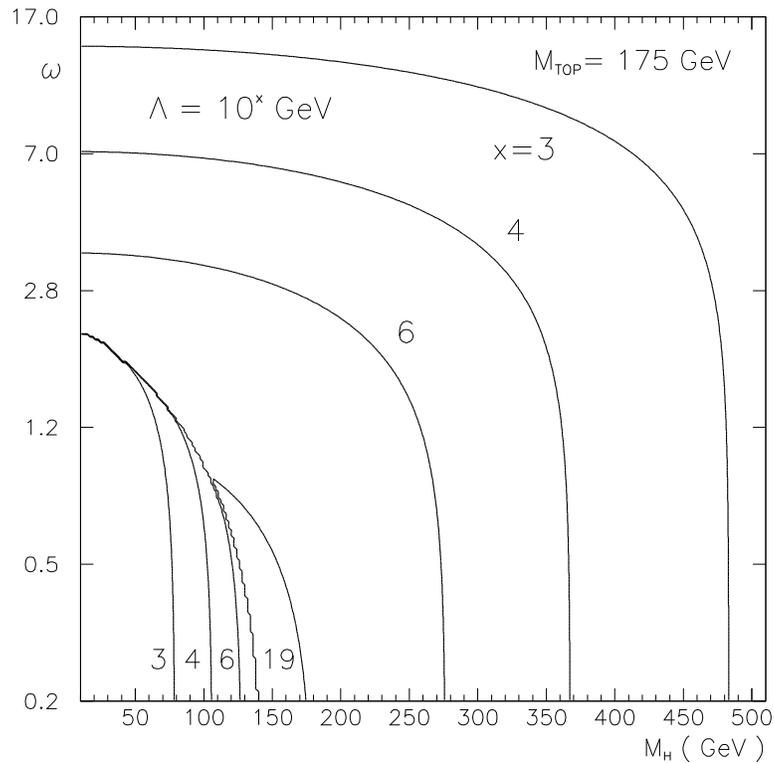,height=10.cm,angle=0}}
\begin{center}
\parbox{13cm}{
\caption{\label{stability} \it Theoretical limits on the parameters of the model
in the $\omega$ vs. $M_H$ plane. The contour lines correspond 
to the cutoff scales $\Lambda = 10^{19}$, $10^6$, $10^4$ and $10^3$ GeV.}}
\end{center}
\end{figure}

At some scale the validity of the one loop RGEs is spoiled by the appearance of 
the Landau-pole. There, some new physics has to appear to cancel the 
occurring divergences. Such a cutoff scale,
$\Lambda$, can be determined, if one defines a condition for a regular behavior
of the theory below that scale. To guarantee regularity, one chooses a 
coupling not to exceed a given value. 
If one of the couplings diverges, all beta functions 
which depend on that coupling
will do so, too. Thus it is enough to impose a cutoff condition for one of the
scalar sector couplings. We choose
the Higgs self coupling. Because one assumes a 
nonperturbative Higgs sector for Higgs masses
($\sim \sqrt{2\lambda}v$) around $700 GeV$, we demand an upper 
bound on $\lambda$.
That condition leads to upper bounds on the couplings of the scalar sector.
On the other hand, because of the large Top Yukawa coupling, the Higgs self 
coupling becomes negative in some parameter space, which signals the 
instability of the vacuum at some scale \cite{sher}. 
Avoiding that below a given scale
gives constraints on the parameters. For the scalar couplings it leads to lower 
bounds. Thus we define a cutoff scale by the requirement 
\begin{equation}
0 < \lambda(t)/(4\pi) < 1/3 \, ,\quad \mbox{for all} \quad t < 
\log(\Lambda/(2M_Z)) \end{equation}  
By calculating the RGEs, Eqn.~\ref{rges}, one obtains 
an allowed parameter region for the initial values of $M_H,\,\omega,\,\kappa$
to a given cutoff scale. By putting $\kappa(2M_Z)=0$, one finds a plot in the 
$(\omega,M_H)$ plane, 
with contours belonging to the chosen scale. Such a plot is given in 
Fig.~(\ref{stability}). 

One finds a region in the parameter space,
where the model is valid up to the Planck energy, $\sim 10^{19}$ GeV.
That is in accordance with lower bounds on the SM Higgs bosons of $\sim 145$ 
GeV found by other authors 
\cite{quiros}. There, the effective potential was used  to derive
a lower bound on the Higgs mass by demanding a perturbative SM up to Planck 
energies. The non-minimal couplings partly cancel the negative contribution 
of the Top Yukawa coupling
in the beta function of the quartic Higgs coupling. This reduces the
lower bounds on the Higgs mass as $\omega$ increases, until the Landau pole 
terminates the validity of the model from above. Increasing the Top mass 
leads to a stronger vacuum 
stability bound but shifts the Landau
pole to higher energies, too. The parameter space, valid for stable models, 
lies somewhat higher, but the allowed area is getting smaller. 
For a Top mass at its lower experimental bound, one finds, that one could have 
a stable model with the possibility of a Higgs boson of mass $>60$ GeV. 
If $M_T=175$ GeV, the lower bound is $M_H=105$ GeV.
In the case of non vanishing $\kappa$, the bounds are more restrictive, 
because the Landau pole moves to lower energies as the coupling increases.

\section{Phenomenological bounds}

\begin{figure}
\centerline{\epsfig{figure=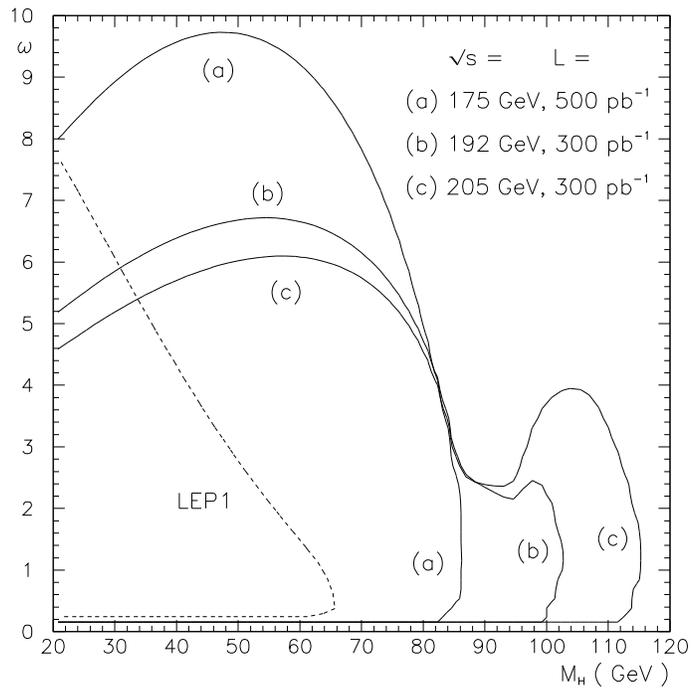,height=9.0cm,angle=0}}
\begin{center}
\parbox{13cm}{
\caption{\label{lepexclu}
\it Exclusion limits at LEP2 (full lines), and LEP1 (dashed). The region
where $\omega$ is small is covered by the search for visible Higgs decay.}}
\end{center}
\end{figure}

We now turn to the phenomenological implications of our model on the Higgs 
search at present and future colliders.
At LEP1 the basic Higgs production mechanism is 
$e^+e^-\rightarrow Z\rightarrow Z^*H^*$
and one looks for the decay products of the off shell bosons. 
For this energy the Higgs boson
mainly decays into a $b$--quark pair. Because of the large QCD background only
the leptonic decays of the Z--boson 
($e^+e^-,\mu^+\mu^-,\nu\bar\nu$) were looked for \cite{opal}.
LEP1 not only provides us with a lower bound on the SM 
Higgs mass of about 60 GeV, 
but also with a bound on invisibly decaying Higgs bosons \cite{nonminh}.
In the case of exclusively invisible decay the mass bound is 65 GeV. 
The bound is higher than in the visible case because one can also look for
hadronic decays of the Z--boson.
We claim, that for fast Higgs decay due to strong nonstandard interactions, 
this bound is not valid. In that case the smearing of the Higgs resonance leads 
to a dilution of the signal to background ratio.  
We calculated the cross section for invisible 
Higgs decay at LEP1 for our model, 
to quantify, how large width effects change the mass bounds. 
The differential cross section for the off shell bosons is given by
\begin{eqnarray}\label{lep1cs}
\frac{d^2\sigma}{ds_I\,ds_Z} &=& \frac{\pi\,(v_e^2+a_e^2)}{192}
\left(\frac{\alpha}{s_W^2c_W^2}\right)^2
\rho_Z(s_Z)\rho_H(s_I) B_{Z\rightarrow f\bar f}(s_Z) 
B_{H\rightarrow \varphi\varphi}(s_I) \nonumber \\&&
\frac{\sqrt{\lambda(M_Z^2,s_Z,s_I)}\,
(\lambda(M_Z^2,s_Z,s_I)+12\,s_Z\, M_Z^2)}{M_Z^4\,\Gamma_Z^2\,s_Z}\\
 && 4m_{\varphi}^2\leq s_H \leq s \, , \quad 0 \leq s_Z \leq 
 (\sqrt{s}-\sqrt{s_H})^2\,, \nonumber
\end{eqnarray}
where we have used the functions
\begin{eqnarray}
 \lambda(x,y,z) &=& x^2+y^2+z^2-2xy-2yz-2zx \nonumber \\
 \rho_X(x)      &=& \frac{1}{\pi}
 \frac{\sqrt{x}\,\Gamma_X(x)}{(x-M_X^2)^2+x\Gamma_X(x)^2} \nonumber 
\end{eqnarray}
The functions $B_X(x)$ are scale dependent branching ratios of the 
respective boson ($X=Z,H$).
The vector and axial vector coefficients are normalized to $a_e=1$.
To get a feeling for the smearing effect at LEP1 we present in 
Tab.~(\ref{lep1tab}) for some values of $\omega/$
the signal cross section for invisible Higgs production with a 
cut in the missing 
invariant mass, $\sqrt{s_I}=M_H\pm 5$ GeV. The Z boson is assumed to
decay into electron, muons and quarks.
\begin{table}[htb]
\begin{center}
\begin{tabular}{|c|c|c|c|c|c|c|}\hline
$\omega\backslash M_H$ & 20    &    30 &    40 &   50 &    60 &   70 \\ \hline
0.0                & 36.31 & 14.68 &  5.57 & 1.81 &  0.44 & 0.058  \\ \hline 
0.1                & 36.00 & 14.51 &  5.47 & 1.76 &  0.44 & 0.055  \\ \hline 
0.3                & 32.47 & 13.51 &  5.24 & 1.72 &  0.42 & 0.055  \\ \hline
0.5                & 23.58 & 11.03 &  4.51 & 1.54 &  0.39 & 0.054  \\ \hline
1.0                &  7.85 &  4.57 &  2.20 & 0.85 &  0.24 & 0.037  \\ \hline
2.0                &  1.97 &  1.20 &  0.61 & 0.25 &  0.07 & 0.012  \\ \hline
4.0                &  0.49 &  0.30 &  0.15 & 0.06 &  0.019 &0.003  \\ \hline
8.0                &  0.12 &  0.07 &  0.04 & 0.02 &  0.005 &0.001  \\ \hline
\end{tabular}
\parbox{13cm}{
\caption{\label{lep1tab}\it Cross section in pb of the  
$ee,\mu\mu,qq+E\!\!\!/$--signal for different
values of $M_H$/GeV.
The numbers are calculated with a cut on the invisible mass of $5$ GeV. 
The first line is the SM result for $ee,\mu\mu,qq+H$.}}
\end{center}
\end{table} 
In Fig.~(\ref{lepexclu}) 
we indicated the excluded region  in the $M_H,\omega$ 
plane for our model by the dashed line. 
Neglecting the background, the exclusion level is defined by the demand, that 
fewer  
than three events in a $\pm 5 \,GeV$ band around the resonance 
maximum in the missing mass
distribution, are present. This is a typical resolution
for a hadronic signal. 
To compare with Ref.~\cite{nonminh}, we assumed a luminosity of 
$50 pb^{-1}$ with an efficiency of $50\%$.
In deriving our bounds, we neglected  the Phion mass, 
thereby obtaining the strongest bound, and the $\kappa$ dependence
which is a small perturbation for not too large values of $\kappa$.

\begin{figure}[htb]
\vspace{0.1cm}
\centerline{\epsfig{figure=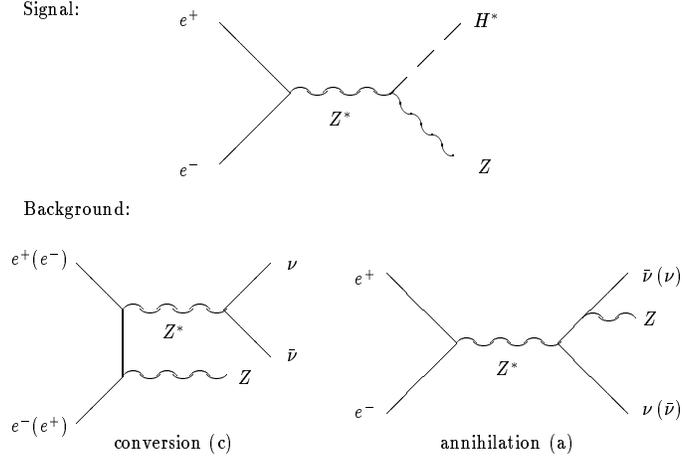,height=6cm,angle=0}}
\begin{center}
\parbox{13cm}{
\caption{\label{lep2graphs}\it Relevant Feynman graphs for 
invisible Higgs search at LEP2.}}
\end{center}
\end{figure}

At LEP2 it should be possible to detect the 
Higgs boson of the SM up to masses of about $(\sqrt{s} - 100)$ GeV 
by the Higgs Bremsstrahlung 
of a virtual Z boson, the Bjorken process \cite{bjo}. 
The Higgs production by $WW$--fusion plays no role at LEP energies.
The cross section has a reasonable size, only if the production of 
an on shell Higgs and Z boson is 
kinematically allowed. An irreducible background stems from
$Z^*Z^*$--production. Assuming the reconstruction of a Z boson, one
has to consider the background reactions $e^+e^- \rightarrow Z f \bar f$, 
where one distinguishes two types of graphs. These are shown together
with the signal graph in Fig.~(\ref{lep2graphs}).

At LEP, the visible Higgs decays inside the SM are heavily suppressed even for
small couplings to the hidden sector. This is due to the 
extreme small Higgs width of a few MeV at LEP energies. 
Even for a narrow invisible Higgs width of about 1 GeV
the branching ratio into Phions is nearly one, as can be deduced from 
Fig. (\ref{width}). Because the visible SM decays are well understood, 
we focus on the invisible ones in the following.
 
\begin{figure}[htb]
\vspace{0.1cm}
\centerline{\epsfig{figure=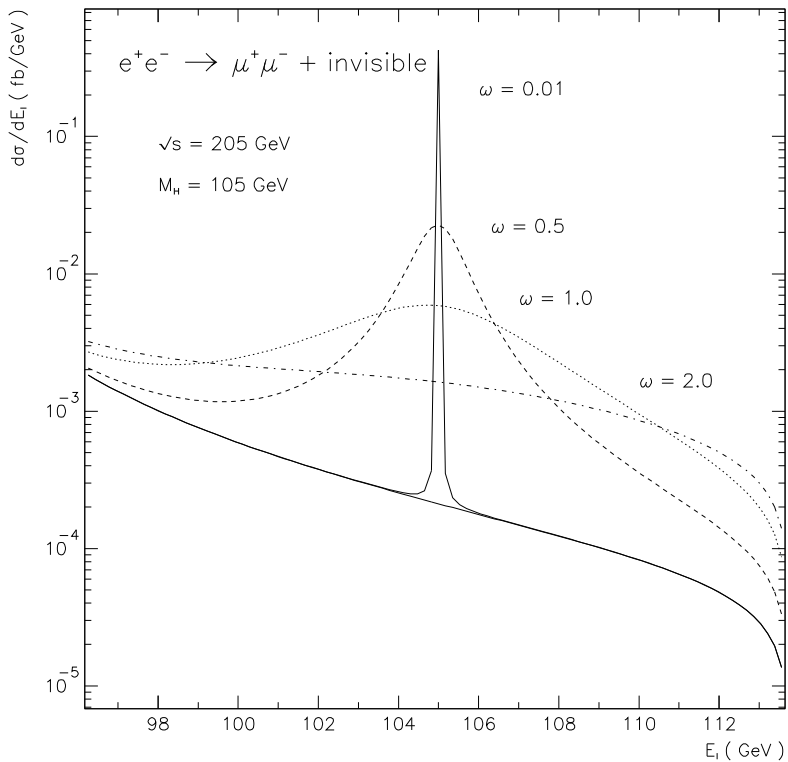,height=10.cm,angle=0}}
\begin{center}
\parbox{13cm}{
\caption[LEP2 missing mass distribution]{\label{sigshape}
\it Example of a missing energy 
differential cross section at LEP2 with $M_H=105$ GeV, $\sqrt{s}=205$ GeV 
for several values of $\omega$.}}
\end{center}
\end{figure}

\begin{figure}[htb]
\vspace{0.1cm}
\centerline{\epsfig{figure=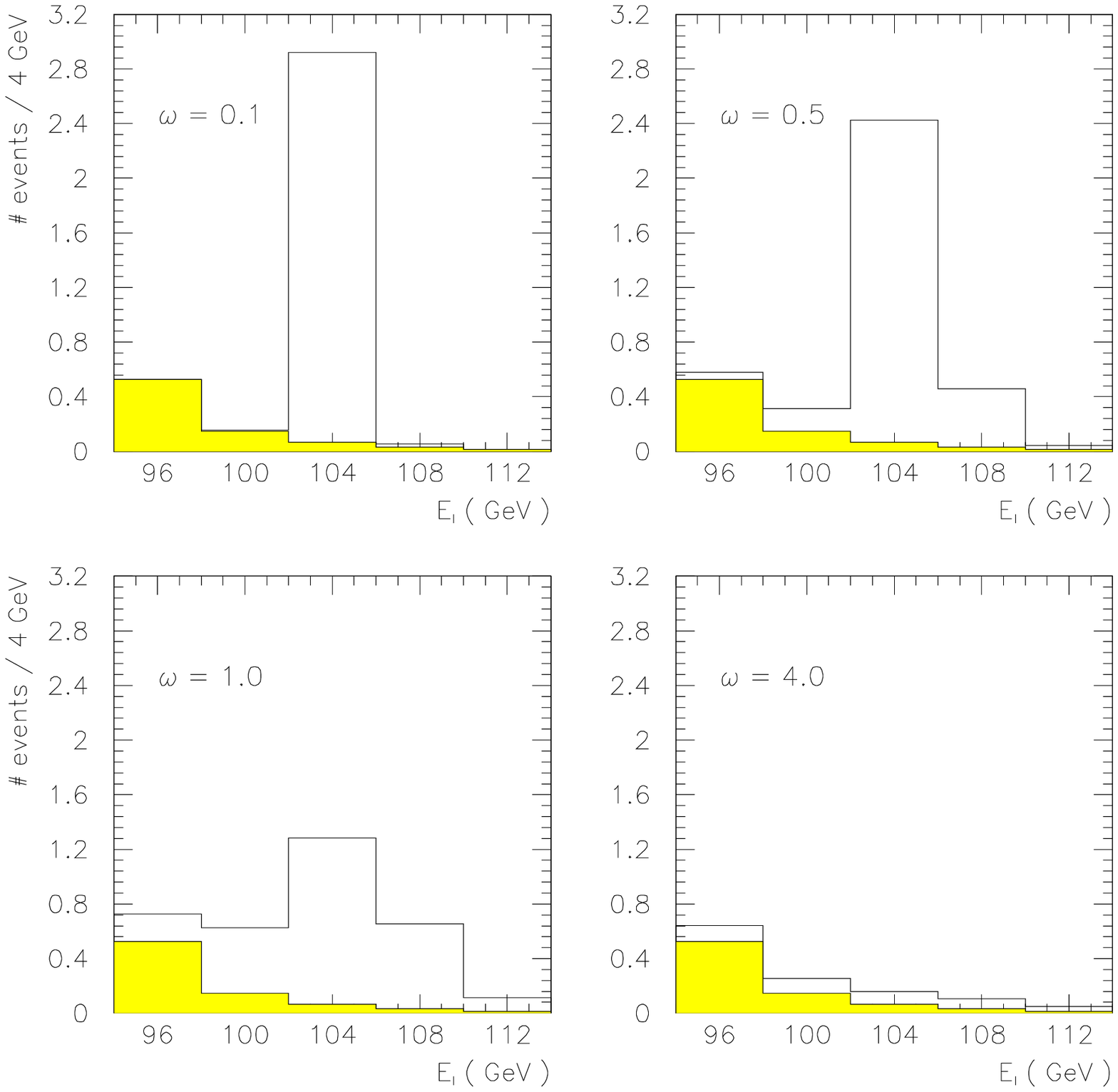,height=10.cm,angle=0}}
\begin{center}
\parbox{13cm}{
\caption[LEP2 event histogram]{\label{eveshape}
\it The event histograms belonging to Fig.~(\ref{sigshape})
assuming  an integrated luminosity of $300 pb^{-1}$. 
The background is shaded.}}
\end{center}
\end{figure}

We first give a list of the used analytic expressions for the $d\sigma/ds_I$,
where $s_I$ is the invariant mass squared of the invisible particles. In the 
formulae the three neutrino species are already taken into account.
\begin{eqnarray} \label{cs}
\frac{d\sigma_H}{ds_I} &=& \frac{\pi(v_e^2+a_e^2)}{192}\left(
\frac{\alpha}{s_W^2c_W^2}\right)^2
\frac{\sqrt{\lambda}\,(\lambda+12\,s\, M_Z^2)}{s^2(M_Z^2-s)^2} \rho_H(s_I)
\\
\frac{d\sigma_{Z^*(c)}}{ds_I} &=& 
\frac{1}{512}(v_e^4+6\,v_e^2a_e^2+a_e^4)(v_f^2+a_f^2)
\left(\frac{\alpha}{s_W^2c_W^2}\right)^3 \nonumber 
\\ && \times \frac{s_I}{s^2\,((M_Z^2-s_I)^2+s_I\Gamma_Z^2)} \,G(s,s_I,M_Z^2)
\\
\frac{d\sigma_{Z^*(a)}}{ds_I} &=& 
\frac{1}{512}(v_f^4+6\,v_f^2a_f^2+a_f^4)(v_e^2+a_e^2)
\left(\frac{\alpha}{s_W^2c_W^2}\right)^3 \nonumber
\\ && \times\frac{1}{s\,(M_Z^2-s)^2} \,G(s_I,s,M_Z^2) 
\\
\end{eqnarray}
The indices for the background cross sections 
correspond to Fig.~(\ref{lep2graphs}). The introduced function is:
\begin{eqnarray}
 G(x,y,z) &=&  \frac{x^2+(y+z)^2}{x-y-z}\,\mbox{arctanh}
 \left(\frac{\sqrt{\lambda(x,y,z)}}{x-y-z}\right)-\sqrt{\lambda(x,y,z)} 
 \nonumber 
\end{eqnarray}
Calculated from the upper formulae, the shape of 
the missing energy distribution 
depending on $\omega$ is shown in Fig.~(\ref{sigshape}) for 
$M_H = 105\,GeV, \sqrt{s}=205\,GeV$ .
Event histograms for the same parameters look like Fig.~(\ref{eveshape}). 
One recognizes that the Higgs peak is smeared out 
considerably, even for not unreasonably high values of $\omega$ 
($<<(16\pi^2)^{-1}$). 
To reduce the numbers of parameters we put $\kappa = 0$ 
and neglected the Phion mass.

To illustrate the dependence of the event rates and cross sections for 
invisible Higgs decay on the non standard
coupling $\omega$ we give the expected number of signal and background events 
for beam energies of $\sqrt{s} = 175,\,192,\,205$ GeV in the following.
It should be noted, that the background is dominated
by the resonating $Z^*$--conversion graph. 
It is considerably reduced by cutting away the 
energy region, where the cross section of the resonating $Z^*$ is big. 
This should be possible, as long as the Higgs peak is sharp and well 
separated from the $Z^*$ resonance.
To define a Higgs signal for a given mass,
we count only events near the resonance maximum, $M_H\pm\Delta$.
We choose $\Delta = 5\,GeV$ in the following, which corresponds to
a typical detector resolution for hadronic energy at LEP. 
The number of events per year related to the signal 
$S$ and background $B$ depend on the expected luminosities:   
\begin{eqnarray} \label{sigback}
S_{\Delta}(\sqrt{s},M_H) &=& L(\sqrt{s})
\int\limits_{(M_H-\Delta)^2}^{(M_H+\Delta)^2} ds_Y 
                             \frac{d\sigma_H}{ds_Y} \\ \nonumber
B_{\Delta}(\sqrt{s},M_H) &=& L(\sqrt{s})
\int\limits_{(M_H-\Delta)^2}^{(M_H+\Delta)^2} ds_Y 
                  \frac{d ( \sigma_{Z(c)} + \sigma_{Z(a)} )}{ds_Y}
\end{eqnarray} 
The assumed LEP2 luminosities are given by L(175 GeV)=500/(pb\,y), 
L(192 GeV) = L(205 GeV) = 300/(pb\,y).

We list the 
number of events, $S$, of  the $\mu\mu+E\!\!\!/$-signal in 
Tab.~(\ref{lep2tab}). Though a resolution
of $4\,GeV$ for the leptonic signal is possible,
we put $\Delta=10$ in the signal definition, Eqn.~\ref{sigback}, 
because we do not want to loose events from a  smeared Higgs
resonance.
To compare these numbers with the hadronic decays of the Z--boson, 
one simply has to multiply with 
$B(Z\rightarrow q\bar q)/B(Z\rightarrow \mu\mu) \simeq 20$.
\begin{table}[htb]
\begin{center}
\begin{tabular}{|r|r|r|r|r|r|r|}  \hline
$\omega$  & $S_{\Delta}(175,70)$ & $S_{\Delta}(192,70)$ & $S_{\Delta}(192,90)$ &
            $S_{\Delta}(205,70)$ & $S_{\Delta}(205,90)$ &
             $S_{\Delta}(205,105)$ \\ \hline
$B_{\Delta}$&\,0.1 (\,0.2)  &  0.1 (1.5) & 4.1 (4.6) &
 0.1 (4.9) & 4.7 (4.9) & 0.4 (4.9) \\ \hline
      .1  &      15.7 (15.7)  &  8.5 (8.5) & 4.7 (4.8) &
       7.3 (7.3) & 5.1 (5.1) & 2.9 (2.9) \\ \hline
      .5  &      15.4 (16.1)  &  8.4 (8.8) & 4.8 (5.0) &
       7.2 (7.5) & 5.2 (5.4) & 3.0 (3.2) \\ \hline
     1.0  &      12.4 (15.3)  &  6.8 (8.5) & 4.1 (4.9) &
      5.8 (7.4) & 4.4 (5.3) & 2.6 (3.1) \\ \hline
     2.0  & \,5.5 (13.5)  &  3.1 (7.6) & 2.1 (4.6) &
      2.6 (6.7) & 2.4 (4.9) & 1.4 (3.0) \\ \hline
     4.0  & \,1.5 (\,5.1)  &  0.8 (2.9) & 0.6 (2.9) &
      0.7 (2.8) & 0.6 (2.8) & 0.4 (2.7) \\ \hline
     8.0  & \,0.4 (\,1.3)  &  0.2 (0.7) & 0.1 (0.7) &
      0.2 (0.7) & 0.2 (0.7) & 0.1 (0.7) \\ \hline 
\end{tabular}
\parbox{13cm}{
\caption[LEP2 cross sections]{\label{lep2tab}
\it Number of events $S_{\Delta}$ of the $\mu\mu+E\!\!\!/$--signal
at LEP2. The first line contents the background.
The event rates are calculated with a cut on the missing mass of 
$\Delta=10$ GeV. In brackets we give
the numbers without  the cut.}}
\end{center}
\end{table}  
Depending on our parameters in the $(M_H,\omega)$--plane we present exclusion 
plots for several center of mass energies in Fig.~(\ref{lepexclu}). 
The $95\%$ confidence level is defined by Poisson statistics \cite{LEP2report}.
The depression for $M_H\sim M_Z$ stems
from the larger background around the Z--resonance of the conversion graph. 
The large Higgs width  leads to a restriction
even for higher Higgs masses than kinematically allowed, 
because one probes the low energy tail of
such a resonance. For a small Higgs width the bound is stronger 
than for a SM Higgs boson, because
one can also use the hadronic decay of the $Z$--boson, 
as was already noticed  for the LEP1 bounds above. 
Again the signal vanishes as the Higgs--Phion coupling
is getting large. 

Fig.~(\ref{lepexclu}) reflects the fact that for a
sufficiently large nonstandard coupling $\omega$, 
Higgs detection at LEP2 may not be possible
even if it is light enough to be produced and its coupling 
to the Z--boson is not reduced
by mixing effects. The bounds tell us the absolute size of nonperturbative
effects from a light hidden sector which would forbid Higgs detection.  

At the NLC the upper limits on the couplings in the present model
come from the invisible decay too, as the branching ratio
into visible particles drops with increasing $\varphi$--Higgs
coupling ($\omega$).
Because above the vector boson threshold the Higgs decay modes into SM particles
are sizeable, as can be seen in Fig.~(\ref{width}), 
one has to consider visible Higgs decays, too.
Since the main source for Higgs production, the $WW$--fusion process,
can not be used to look for invisible Higgs decay,
one is in principle left with the Higgs-Strahlung and $ZZ$--fusion reaction. 
For energies up to 500 GeV the Higgs-Strahlungs cross section is dominant and
is of comparable size as the $ZZ$--fusion process even if one is folding in 
the branching ratio $B(Z\rightarrow e^+e^-,\mu^+\mu^-)$. 
The possibility to tag an on--shell Z boson via a leptonic 
system which is extremely useful 
for the discrimination of possible backgrounds makes Higgs-strahlung 
to be the preferred production
mechanism. Thus we only have considered reactions 
containing an on shell Z boson with its decay into $e^+e^-$ or $\mu^+\mu^-$.  
One should be aware that a few events from the huge 
$WW$ background may survive, but
that the $Z\nu\nu$ background is dominant after imposing the cuts defined below.
A detailed discussion on possible backgrounds can be found in Ref.~\cite{eboli}.
Then the signal cross section is the well  known 
Higgs-Strahlungs cross section, Eqn. (\ref{cs}), modified
by the non standard Higgs width due to Phion decay.
We calculated the $Z\nu\nu$ background with the relevant set of graphs for
Z production ($ZZ$--production, $WW$--fusion and Z initial, 
final state radiation) by a
Monte Carlo program. The calculated amplitudes are 
identical to the result of Ref.~\cite{mele}. 
To reduce the background we have used the fact
that the angular distribution of the Z--boson 
for the signal has a peak for small values
of $|\cos\theta_Z|$ in contrast to the background. 
Thus we imposed an angular cut $|\cos\theta_Z|<0.7$.
Because we assume the reconstruction of the on--shell 
Z--boson we use the kinematical relation
$E_Z=(s+M_Z^2-s_I)/(2\sqrt{s})$
between the Z energy and the invariant mass of the invisible system
to define a second cut. Since the differential cross section 
$d\sigma/ds_I$ contains the 
Higgs resonance at $s_I=M_H^2$, we impose the following condition on the Z 
energy
\begin{equation} \frac{s+M_Z^2-(M_H+\Delta)^2}{2\sqrt{s}}<E_Z<
\frac{s+M_Z^2-(M_H-\Delta)^2}{2\sqrt{s}}
\end{equation} 
which is equivalent to a cut on the invisible mass.
As long as the Higgs width is small, one
is allowed to use small  $\Delta$, which reduces the background 
considerably keeping most of the signal events. But in the case of large 
$\varphi$--Higgs coupling, $\omega$, one
looses valuable events. To compromise between both effects we took  
$\Delta=30$ GeV. Some cross sections are given in Tab.~(\ref{nlccs}). 
\begin{table}[htb]
\begin{center}
\begin{tabular}{|r|r|r|r|r|r|r|}  \hline
$\omega/M_H$    &   150 &   200 &  250 &   300 &   350 &   400  \\ \hline
$B_{\Delta}$    &   2.51  &   4.73  &  10.01 &   21.44  &  44.85  &  39.10   \\ 
\hline
      .1 &     31.90  &  25.24  &  18.02  &  11.12 &    5.49  &   1.16    \\ 
      \hline
      .3 &     31.89  &  25.11  &  17.98  &  11.12 &    5.48  &   1.16    \\ 
      \hline
      .5 &     31.57  &  25.03  &  17.88  &  11.06 &    5.45  &   1.15    \\ 
      \hline
     1.0 &     30.58  &  24.40  &  17.56  &  10.91 &    5.40  &   1.14    \\ 
     \hline
     2.0 &     26.32  &  21.98  &  16.16  &  10.17 &    5.09  &   1.10    \\ 
     \hline
     4.0 &     15.09  &  14.27  &  11.41  &   7.66 &    4.00  &    .93    \\ 
     \hline
     8.0 &      4.61  &   4.85  &   4.32  &   3.22 &    1.88  &    .50    \\ 
     \hline 
    16.0 &      1.17  &   1.25  &   1.15  &    .89 &     .55  &    .16    \\ 
    \hline
\end{tabular}
\parbox{13cm}{
\caption[NLC cross sections]{\label{nlccs}
\it Cross sections of the  $ee,\mu\mu,qq+E\!\!\!/$--signal
at the NLC with $\sqrt{s}=500$ GeV in fb. $M_H$ is given in GeV.
The numbers are calculated with cuts on the invisible mass, $M_H\pm30$ GeV, and 
polar angle of the Z--boson, $|\cos(\theta_Z)|\leq 0.7$ . The first
line shows the background due to $Z\nu\nu$ reactions.}}
\end{center}
\end{table}  

\begin{figure}[htb]
\vspace{0.1cm}
\centerline{\epsfig{figure=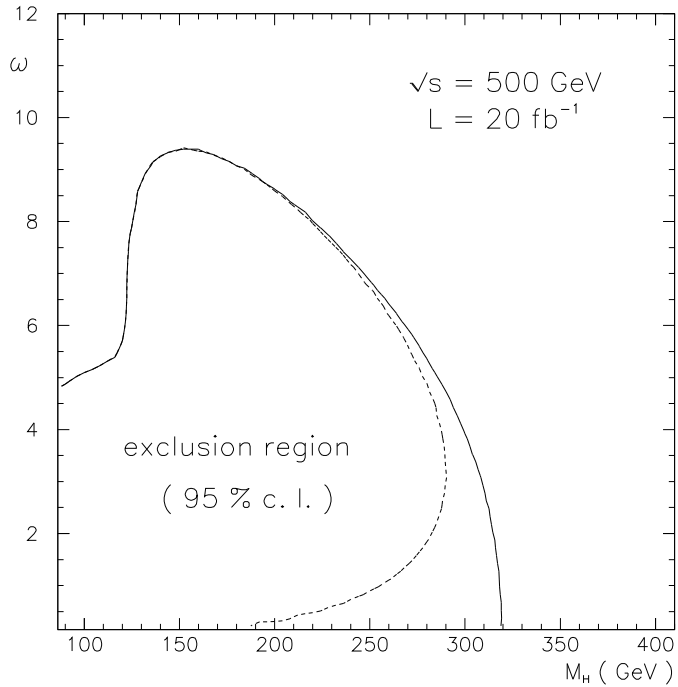,height=9.cm,angle=0}}
\begin{center}
\parbox{13cm}{
\caption{\label{nlcexclu}
\it Exclusion limits at the NLC due to Higgs searches. The dashed
line corresponds to the invisible, the full line to all Higgs decay modes.}}
\end{center}
\end{figure}

For the exclusion limits we assumed an integrated luminosity
of $20\,fb^{-1}$. 
The result is given  
in Fig.~\ref{nlcexclu}. Again, one notices the somewhat reduced sensitivity
for the mass region where $M_H\simeq M_Z$. For larger
values of $M_H$ the limit stems from the other $Z\nu\nu$ backgrounds
with $W$ bosons in the t--channel and from kinematical constraints. 
For large $\omega$ the signal ceases
to dominate over the background because the Higgs peak is smeared out
to an almost flat distribution. 

To end this section, we want to comment on invisible Higgs search at LHC.
Below the two Z boson threshold the Bjorken process is the main production
mechanism for Higgs bosons, too. But because of the unknown longitudinal
momentum of the partons, the missing mass is not an observable. Without the 
possibility of reducing the background through an adequate cut, one is not
able to exploit the kinematical potential of the collider. The authors 
of Refs.~\cite{lhcsearch} give upper mass bounds for the 
detectability of an invisible
decaying Higgs boson between 150 and 250 GeV assuming a high yearly 
luminosity of $100 fb^{-1}$. For higher masses the cross section is simply
to small. The most promising production mechanism for Higgs bosons with
masses beyond the ZZ threshold, Higgs production by gluon fusion, 
cannot be used here due to the suppression of 
the branching ratio $B(H\rightarrow visible)$ in our model
which is a small number in the case of strong Higgs--Phion 
interactions \cite{vladimir}.  
We see that the LHC is not competitive to
$e^+e^-$ colliders for invisible Higgs search.

\section{Conclusions}
To understand and quantify the effect of light hidden matter on Higgs 
search at present and future colliders
we introduced  a scalar sector coupled to the SM Higgs sector. 
Explicit formulae for the modified Higgs-propagator were presented.
We have shown that the main effect is that the Higgs width is becoming 
a free parameter. Our model is constrained by 
theoretical and phenomenological bounds on the parameters.
First we determined the allowed region of the nonstandard couplings 
in respect to vacuum instability and the Landau pole by analyzing the 
RGEs of our model. Then we have discussed, how present experimental 
LEP1 data is  restricting the model parameters 
already and how these bounds will be increased by LEP2 and NLC.
We pointed out, that at LHC the presented bounds can not be extended. 
This is mainly
because the invisible mass is not an observable. Comparing all
results, we conclude that restrictive bounds on the a parameter region can be
obtained by analyzing missing energy signals. But for 
sufficiently large coupling
between the Higgs boson and the hidden sector particles the Higgs boson 
could escape detection, even if it is light enough to be produced.
If no evidence for a SM Higgs boson occurs at these colliders, either
by direct measurement of a Higgs below $\sim 700$ GeV, 
or through some Higgs
remnant  resonance in $W_L W_L$-scattering in the TeV range, strongly
coupled light hidden matter could be the reason.

\end{document}